\documentclass[12pt]{article}
\usepackage{graphicx}
\date{}
\def\be{\begin{equation}}
\def\ee{\end{equation}}
\def\bea{\begin{eqnarray}}
\def\eea{\end{eqnarray}}
\def\s{\sigma}
\def\al{\alpha}

\def\om{\omega}
\def\pr{\prime}

\def\th{\theta}
\def\ep{\varepsilon}

\title{STRING MODELS OF GLUEBALL \\
AND REGGE TRAJECTORIES \thanks{Is supported
by Russian foundation of basic research, grant 05-02-16722}\\
}

\author{G.\,S. Sharov\\
{\small Tver state university}\\
{\small Tver, 170002, Sadovyj per. 35, Mathem. dep-t}}
\begin{document}
\maketitle
\begin{abstract}
The closed relativistic string carrying two point-like masses is
considered as the model of a glueball with two constituent gluons.
Here the gluon-gluon interaction is simulated by a pair of strings.
For this system exact solutions of classical equations of motions
are obtained. They describe rotational states of the string
resulting in the set of quasilinear Regge trajectories with
different behavior.
\end{abstract}

\centerline {\bf Introduction}
\medskip

In various string models of mesons and baryons
\cite{Nambu}\,--\,\cite{Solovm} the Nambu-Goto string simulates
strong interaction between quarks at large distances and the QCD
confinement mechanism. This string has linearly growing energy
(energy density is equal to the string tension $\gamma$) and
accounts for the nonperturbative contribution of the gluon field.

All cited string hadron models generate linear or quasilinear Regge
trajectories
 \be
 J\simeq\al_0+\al'M^2,
 \label{Reggm} \ee
 where $J$ and $M$ are the angular momentum and energy of a hadron state,
the slope $\al'\simeq0{.}9$ GeV$^{-2}$ for the simplest model of
meson is $\al'=1/(2\pi\gamma)$. The string models describe
excited hadron states on the leading Regge trajectories if we use
the rotational states of these systems (planar uniform rotations)
for this purpose.

These properties of the string were used for describing glueballs
(bound states of gluons) on the base of different string models
\cite{Yamada}\,--\,\cite{ZayasS}, including combinations of string
and potential approaches \cite{BrauS}\,--\,\cite{MathieuS3}. The
glueballs are predicted in QCD and observed in lattice QCD
simulations \cite{MorningP,LiuWu,MeyerT}.

The QCD glueball may be identified with the pomeron that is the
Regge pole determining an asymptotic behavior of high-energy
diffractive amplitudes \cite{DonPom,Kaidal01}. The pomeron Regge
trajectory \cite{MeyerT,DonPom}
 \be
 J\simeq1{.}08+0{.}25M^2
 \label{ReggPom} \ee
 differs from hadronic ones (\ref{Reggm}), in particular,
its slope is essentially lower.

String models of glueballs describing trajectories of the type
(\ref{ReggPom}) or some exotic hadron states (glueball candidates)
were suggested in the following variants
\cite{Yamada}\,--\,\cite{MathieuS}: (a) the open string with
enhanced tension (the adjoint string) and two constintuent gluons at
the endpoints; (b) the closed string simulating gluonic field; (c)
the closed string carrying two point-like masses (constituent
gluons). Evidently, the last model may be easily generalized for
three-gluon glueballs \cite{MathieuS3}.

In this paper the string model (c) (closed string with two masses
$m_1$ and $m_2$) is considered, but it includes as particular cases
the model (a) (if $m_1=m_2=0$) and the closed string
carrying one massive point ($m_2=0$) \cite{MilSh}. The last model
may describe gluelumps \cite{AbreuB}, \cite{KarlP}.

In Sect.~1 we describe the classical dynamics of the considered
string model (c) moving in Minkowski space $R^{1,3}$ and in its
generalization --- in the space ${\cal M}=R^{1,3}\times T^{D-4}$
\cite{MilSh}. Here $T^{D-4}=S^1\times S^1\times\dots\times S^1$ is
$D-4$\,-\,dimensional torus with flat metric resulting from the
compactification procedure of the string theory \cite{GSW}. The
space ${\cal M}$ has nontrivial structure of its homotopic classes.
It is essential for states of a closed string. In the particular
case $D=4$ we simply have ${\cal M}=R^{1,3}$, so this case is also
included into consideration.

In Sect.~2 rotational motions (planar uniform rotations) of this
system are described and classified. They have much more complicated
structure than a well known set of rotations of the folded
rectilinear string. Rotational motions of string systems are widely
used for generating Regge trajectories. Their structure and behavior
for the considered system are described in Sect.~3.

\bigskip

\centerline {\bf 1. Dynamics}
\medskip

The dynamics of the closed string carrying two point-like masses
$m_1$ and $m_2$ is determined by the action close to that of the
string baryon model ``triangle" \cite{PRTr}:
 \be
S=-\gamma\int\limits_{\Omega}\sqrt{-g}\;d\tau d\s
-\sum\limits_{i=1}^2m_i\int\sqrt{\dot x_i^2(\tau)}\;d\tau.
 \label{S}\ee
 Here $\gamma$ is the string tension, $g$ is the determinant of the
induced metric $g_{ab}=\eta_{\mu\nu}\partial_aX^\mu\partial_bX^\nu$
on the string world surface $X^\mu(\tau,\s)$, embedded in ${\cal
M}=R^{1,3}\times T^{D-4}$ (including the particular case $D=4$,
${\cal M}=R^{1,3}$), the speed of light $c=1$. The world surface
mapping in  ${\cal M}$ from
$\Omega=\{\tau,\s:\,\tau_1<\tau<\tau_2,\,\s_0(\tau)<\s<\s_2(\tau)\}$
is divided by the world lines of massive points
$x_i^\mu(\tau)=X^\mu(\tau,\s_i(\tau))$, $i=0,1,2$ into two world
sheets. Two of these functions $x_0$ and $x_2$ describe the same
trajectory of the 2-nd massive point, and their equality forms the
closure condition
 \be
X^\mu(\tau,\s_1(\tau))\stackrel{\sim}{=}X^\mu(\tau^*,\s_2(\tau^*))
\label{close}\ee
 on the tube-like world surface \cite{PRTr,MilSh}. These equations may
contain two different parameters $\tau$ and $\tau^*$, connected via
the relation $\tau^*=\tau^*(\tau)$. This relation should be included
in the closure condition (\ref{close}) of the world surface.

The symbol of equality $\stackrel{\sim}{=}$ in the condition
(\ref{close}) means usual equality for coordinates $x^0$, $x^1$,
$x^2$, $x^3$ in $R^{1,3}$, but for other coordinates $x^k$,
$k=4,5\dots$ (describing the torus $T^{D-4}$) this includes their
cyclicity. Namely, points marked by $x^k$ and $x^k+N_k\ell_k$,
$N_k\in Z$ are identified: $x^k\stackrel{\sim}{=}x^k+N\ell_k$.

Equations of motion of this system result from the action (\ref{S})
and its variation. They may be reduced to the simplest form under
the orthonormality conditions on the world surface
 \be
(\partial_\tau X\pm\partial_\s X)^2=0,
 \label{ort}\ee
 and the conditions
 \be \s_0(\tau)=0,\qquad \s_2(\tau)=2\pi.
\label{ends}\ee
 Conditions (\ref{ort}), (\ref{ends}) always may be
fixed without loss of generality, if we choose the relevant
coordinates $\tau$, $\s$ \cite{PRTr}. It is connected with the
invariance of the action (\ref{S}) with respect to nondegenerate
reparametrizations on the world surface
$\tau=\tau(\tilde\tau,\tilde\s)$, $\s=\s(\tilde\tau,\tilde\s)$. The
scalar square in Eq.~(\ref{ort}) results from scalar product
$(\xi,\zeta)=\eta_{\mu\nu}\xi^\mu\zeta^\nu$.

The orthonormality conditions (\ref{ort}) are equivalent to the
conformal flatness of the induced metric $g_{ab}$. Under conditions
(\ref{ort}), (\ref{ends}) the string motion equations take the form
 \cite{PRTr,4B,MilSh}
 \be \frac{\partial^2X^\mu}{\partial\tau^2}-
\frac{\partial^2X^\mu}{\partial\s^2}=0,
 \label{eq}\ee
\vspace{-3mm}
 \be m_1\frac d{d\tau}\frac{\dot
x_1^\mu(\tau)}{\sqrt{\dot x_1^2(\tau)}}+\gamma
\Big[X^{'\!\mu}+\dot\s_1(\tau)\dot
X^\mu\Big]\Big|_{\s=\s_1-0}-\gamma\Big[X^{'\!\mu}+\dot\s_1(\tau)\dot
X^\mu\Big]\Big|_{\s=\s_1+0}=0.
 \label{qq1}\ee
 \vspace{-3mm}
 \be m_2\frac d{d\tau}\frac{\dot x_2^\mu(\tau)}{\sqrt{\dot
x_2^2(\tau)}}+\gamma
\Big[X^{'\!\mu}\big(\tau^*(\tau),2\pi\big)-X^{'\!\mu}(\tau,0)\Big]=0.
\label{qq2}\ee
 Here $\dot X^\mu\equiv\partial_\tau X^\mu$,
$X^{'\!\mu}\equiv\partial_\s X^\mu$.

Eqs.~(\ref{qq1}), (\ref{qq2}) are equations of motion for the
massive points resulting from the action (\ref{S}). They may be
interpreted as boundary conditions for Eq.~(\ref{eq}).

We denote the unit vectors
 $$ e^\mu_0,\;e^\mu_1,\;e^\mu_2,\,\dots\,
e^\mu_{D-1},$$
 associated with coordinates $x^\mu$.
These vectors form the orthonormal basis in the manifold ${\cal M}$.

The closure condition (\ref{close}) under equality  (\ref{ends}) and
the mentioned cyclicity of coordinates $x^k$, $k>3$ takes the form
 \be X^\mu(\tau^*,2\pi)=X^\mu(\tau,0)+\sum\limits_{k\ge4}N_k\ell_k
e^\mu_k \label{clos}\ee

The system of equations (\ref{ort})\,--\,\ref{clos}) describe
dynamics of the closed string carrying two point-like masses without
loss of generality. One also should add that the tube-like world
surface of the closed string is continuous one, but its derivatives
may have discontinuities at the world lines of the massive points
(except for derivatives along these lines) \cite{PRTr}. These
discontinuities are taken into account in Eq.~(\ref{qq1}).

\bigskip

\centerline {\bf 2. Rotational states}
\medskip

We search rotational solutions of system
(\ref{ort})\,--\,\ref{clos}) using the approach supposed in
Ref.~\cite{PRTr} for the string model ``triangle'' and in
Ref.~\cite{MilSh} for the closed string carrying one massive point.
In the frameworks of the orthonormality gauge (\ref{ort}) we suppose
that the system uniformly rotates, masses move at constant speeds
$v_1$ and $v_2$ along circles and conditions
 \be
\s_1(\tau)=\s_1={}\mbox{const},
 \label{s1}\ee
\vspace{-3mm}
 \be
 \tau^*=\tau+\tau_0,\qquad\tau_0\equiv2\pi\th={}\mbox{const},
 \label{tau0}\ee
 and $\dot X^2(\tau,\s_i)={}\mbox{const}$ or
 \be
\frac\gamma{m_1}\sqrt{\dot X^2(\tau,\s_1)}=Q_1,\qquad
\frac\gamma{m_2}\sqrt{\dot X^2(\tau,0)}=Q_2,\qquad
Q_i={}\mbox{const},
 \label{Qi}\ee
 are fulfilled.

When we search solution of the linearized system
(\ref{ort})\,--\,(\ref{clos}) under restrictions
(\ref{s1})\,--\,(\ref{Qi}) as a linear combination of terms
$X^\mu(\tau,\s)= T^\mu(\tau)\,u(\s)$ (Fourier method) we obtain from
Eq.~(\ref{eq}) two equations for functions $T^\mu(\tau)$ and
$u(\s)$:
 $$
T_\mu^{\pr\pr}(\tau)+\om^2 T_\mu=0,\qquad  u''(\s)+\om^2 u=0.$$

Their solutions describing uniform rotations of the string system
(rotational states) contain one nonzero frequency $\om$ and have the
following form \cite{PRTr,MilSh}:
 \be
X^\mu(\tau,\s)=x_0^\mu+e_0^\mu(a_0\tau+b_0\s)+
\sum\limits_{k\ge3}e^\mu_k(a_k\tau+b_k\s)+u(\s)\cdot
e^\mu(\om\tau)+\tilde u(\s)\cdot\acute e^\mu(\om\tau).
 \label{X}\ee

Here
 $$e^\mu(\om\tau)=e^\mu_1\cos\om\tau+e^\mu_2\sin\om\tau,\qquad
\acute e^\mu(\om\tau)=-e^\mu_1\sin\om\tau+e^\mu_2\cos\om\tau$$
 are unit orthogonal vectors rotating in the plane $e_1,\,e_2$;
the functions
 $$
u(\s)=\left\{\begin{array}{ll}A_1\cos\om\s+B_1\sin\om\s,&
\s\in[0,\s_1],\\
A_2\cos\om\s+B_2\sin\om\s,&\s\in[\s_1,2\pi],\end{array}\right.\;\;
\tilde u(\s)=\left\{\begin{array}{ll}\tilde A_1\cos\om\s+\tilde
B_1\sin\om\s,&
\s\in[0,\s_1],\\
\tilde A_2\cos\om\s+\tilde B_2\sin\om\s,&\s\in[\s_1,2\pi],
\end{array}\right.
 $$
 are continuous, but their derivatives
 have discontinuities at $\s=\s_1$ (the position of mass $m_1$).

Continuity of functions $u(\s)$ and $\tilde u(\s)$ at $\s=\s_1$
results in equalities
 \be
A_1C_1+B_1S_1=A_2C_1+B_2S_1,\qquad \tilde A_1C_1+\tilde B_1S_1
=\tilde A_2C_1+\tilde B_2S_1.
 \label{A1}\ee

Here and below we use the notations
 \be
\begin{array}{c}C_1=\cos\om\s_1,\\ S_1=\sin\om\s_1,\end{array}\quad
\begin{array}{c}C =\cos2\pi\om, \\ S=\sin2\pi\om,\end{array}\quad
\begin{array}{c}C_2=\cos(2\pi-\s_1)\om,\\ S_2=\sin(2\pi-\s_1)\om,\end{array}\quad
\begin{array}{c}C_\th =\cos2\pi\th\om,\\ S_\th=\sin2\pi\th\om.\end{array}
 \label{CS} \ee

Expression (\ref{X}) is the solution of Eq.\,(\ref{eq}) and it must
satisfy the conditions (\ref{ort}), (\ref{qq1}), (\ref{qq2}),
(\ref{clos}) under restrictions (\ref{s1})\,--\,(\ref{Qi}). Boundary
condition (\ref{qq1}) with adding Eq.~(\ref{Qi}) takes the form
 $$
\ddot
X(\tau,\s_1)+Q_1\Big[X^{'\!\mu}(\tau,\s_1-0)-X^{'\!\mu}(\tau,\s_1+0)\Big]=0.
 $$
 Substituting Eq.\,(\ref{X}) into this relation we obtain the
equation
 \be
A_2S_1-B_2C_1=A_1(S_1+h_1C_1)+B_1(h_1S_1-C_1)
 \label{A2}\ee
 and the same relation for $\tilde A_i$, $\tilde B_i$. Here we
denote the constants
 \be
h_1=\frac\om{Q_1},\qquad h_2=\frac\om{Q_2}.
 \label{h12}\ee

From the system (\ref{A1}), (\ref{A2}) one can express the
amplitudes $A_2$, $B_2$ via $A_1$, $B_1$. The expressions for
$\tilde A_2$, $\tilde B_2$ are the same, so it is convenient to use
the matrix notations
$${\cal A}=\left(\begin{array}{c} A_1\\ \tilde A_1\end{array}\right),\qquad
{\cal B}=\left(\begin{array}{c} B_1\\ \tilde B_1\end{array}\right),
$$ and present these expressions in the form
 \be
\left(\begin{array}{c} A_2\\ \tilde A_2\end{array}\right)=
(1+h_1C_1S_1)\,{\cal A}+h_1S_1^2{\cal B},\qquad
\left(\begin{array}{c} B_2\\ \tilde B_2\end{array}\right)=
-h_1C_1^2{\cal A}+(1-h_1C_1S_1)\,{\cal B}.
 \label{A2B2}\ee

Substituting expression (\ref{X}) into the closure condition
(\ref{clos}) and taking into account Eqs.~(\ref{A2B2}) we obtain the
following relations for amplitudes:
 \be
b_0=-\th a_0,\qquad b_3=-\th a_3,\qquad b_k=-\th
a_k+\frac{\ell_kN_k}{2\pi}\quad (k>3),
 \label{bak}\ee
\vspace{-2mm}
 \be
M_\th\Big[(C-h_1C_1S_2)\,{\cal A}+(S-h_1S_1S_2)\,{\cal B}\Big]=
{\cal A}
 \label{M1}\ee
 In the latter matrix equality we use the matrix
 $$
M_\th=\left(\begin{array}{cc} C_\th & -S_\th\\ S_\th & C_\th
\end{array}\right)$$
 and notations (\ref{CS}).

Boundary condition (\ref{qq2}) with Eqs.~(\ref{tau0}), (\ref{Qi})
for the solution (\ref{X}) reduces into the matrix equation
 \be
M_\th\Big[-(S+h_1C_1C_2)\,{\cal A}+(C-h_1S_1C_2)\,{\cal B}\Big]=
h_2{\cal A}+{\cal B}.
 \label{M2}\ee

The system of matrix equations (\ref{M1}), (\ref{M2}) may be
rewritten in the form
 \be M_1{\cal A}=M_2{\cal B},\qquad M_3{\cal
A}=M_4{\cal B},
 \label{M14}\ee
\vspace{-2mm}

\noindent where matrices

 \vspace{-2mm}
$$\begin{array}{c}
M_1=(C-h_1C_1S_2)\,M_\th-I,\quad
M_2=-(S-h_1S_1S_2)\,M_\th,\\
M_3=(S+h_1C_1C_2)\,M_\th+h_2I,\quad M_4=(C-h_1S_1C_2)\,M_\th-I
\end{array}$$
 are linear combinations of $M_\th$ and the identity matrix $I$.

Taking into account mutual commutativity of the matrices $M_k$ and
excluding the column ${\cal B}$ (or ${\cal A}$) from the system
(\ref{M14}) we obtain the system equivalent to Eqs.~(\ref{M14})
 \be
M{\cal A}=0,\qquad M{\cal B}=0.
 \label{M0}\ee
 Here the matrix $M=M_1M_4-M_2M_3$ may be reduced with using Eqs.~(\ref{CS})
and equality $M_\th^2=2C_\th M_\th-I$ to the following form:
 $$M=\Big[2(C_\th-C)+(h_1+h_2)\,S-h_1h_2S_1S_2\Big]M_\th.$$

The system (\ref{M0}) (or (\ref{M14})) has nontrivial solutions if
and only if $\det M=0$ that is
 \be
2(C_\th-C)+(h_1+h_2)\,S-h_1h_2S_1S_2=0.
 \label{eqC}\ee

This equation may be rewritten after expanding notations (\ref{CS}),
(\ref{h12}) in the form
 $$2(\cos2\pi\th\om-\cos2\pi\om)+\frac{Q_1+Q_2}{Q_1Q_2}\om\sin2\pi\om=
\frac{\om^2}{Q_1Q_2}\sin\s_1\om\sin d_2\om,$$
 where $d_2=\s_2-\s_1=2\pi-\s_1$. It connects unknown (for the present moment)
values of parameters $\om$, $\th$, $\s_1$, $Q_i$.

Other relations connecting these parameters we obtain after
substituting expression (\ref{X}) into the orthonormality conditions
(\ref{ort}):
 \bea
 &\om^2(A_i^2+B_i^2+\tilde A_i^2+\tilde B_i^2)=
 a_0^2(1+\th^2)-\sum\limits_{k\ge3}(a_k^2+b_k^2),
\qquad i=1,2;&\label{ABort1}\\
&\om^2(\tilde A_iB_i-A_i\tilde B_i)= a_0^2\th+
\sum\limits_{k\ge3}a_kb_k,\qquad i=1,2.&
 \label{ABort2}\eea

Among two equations (\ref{ABort2}) only one is independent, for
example, with $i=1$. If it's satisfied and the relations
(\ref{A2B2}) take place --- the second conditions (\ref{ABort2}) is
satisfied too. But two equations (\ref{ABort1}) are independent.
Below we use the first of them and their residual
 \be
C_1(h_1C_1+2S_1)(A_1^2+\tilde A_1^2)+S_1(h_1S_1-2C_1) (B_1^2+\tilde
B_1^2)=2(C_1^2-S_1^2-h_1C_1S_1)(A_1B_1+\tilde A_1\tilde B_1).
 \label{ABort3}\ee
 Here Eqs.~(\ref{A2B2}) are used.

Under condition (\ref{eqC}) the matrix $M=0$ in Eq.~(\ref{M0}) and
an arbitrary nonzero column ${\cal A}$ or ${\cal B}$ is its
eigenvector. It is connected with the rotational symmetry of the
problem. So one can choose an optional pair $A_1$\,\&\,$\tilde A_1$,
$B_1$\,\&\,$\tilde B_1$ or $A_1$\,\&\,$B_1$ and determine two other
constants from Eqs.~(\ref{M14}) (under condition (\ref{eqC}) two
matrix equations (\ref{M14}) are equivalent), in particular:
 \be
\tilde A_1=\frac{C_*A_1+S_*B_1}{S_\th},\qquad \tilde
B_1=-\frac{KA_1+C_*B_1}{S_\th}.
 \label{tilAB}\ee
 Here the coefficients
 $$C_*=C-h_1C_1S_2-C_\th,\qquad
S_*=S-h_1S_1S_2,\qquad K=S+h_1C_1C_2+h_2C-h_1h_2C_1S_2$$
 are connected by the following relation resulting from
Eq.~(\ref{eqC}):
 \be
C_*^2+S_\th^2=KS_*.
 \label{CSK}\ee

Values (\ref{tilAB}) must obey
Eqs.~(\ref{ABort1})\,--\,(\ref{ABort3}) descending from the
orthonormality conditions (\ref{ort}). If we substitute relations
(\ref{tilAB}) into the first two equations and use Eqs.~(\ref{eqC}),
(\ref{CSK}), we obtain correspondingly
 \be
\frac{\om^2A_*^2}{S_\th^2}\Big[2S+(h_1+h_2)\,C-h_1h_2C_1S_2\Big]=
a_0^2(1+\th^2)-\sum\limits_{k\ge3}(a_k^2+b_k^2),
 \label{a01}\ee
\vspace{-1ex}
 \be
\frac{\om^2A_*^2}{S_\th}= a_0^2\th+\sum\limits_{k\ge3}a_kb_k.
 \label{a02}\ee
 Here the amplitude factor
 \be
A_*^2=KA_1^2+2C_*A_1B_1+S_*B_1^2.
 \label{AA}\ee

If we substitute relations (\ref{tilAB}) into Eq.~(\ref{ABort3}) it
transforms into identity.

The solution (\ref{X}) must obey the last restrictions (\ref{Qi}).
This fact and Eqs.~(\ref{tilAB}), (\ref{CSK}) result in relations
 \be
\begin{array}{c}
a_0^2-\sum\limits\limits_{k\ge3}a_k^2-\om^2A_*^2(S-h_2S_1S_2)\,S_\th^{-2}
=m_1^2Q_1^2\big/\gamma^2,\\
a_0^2-\sum\limits\limits_{k\ge3}a_k^2-\om^2A_*^2(S-h_1S_1S_2)\,S_\th^{-2}
=m_2^2Q_2^2\big/\gamma^2.\rule{0mm}{5mm}
\end{array}
 \label{a0Q}\ee

Below we suppose
 \be
a_k=0\quad\mbox{for}\quad k\ge3
 \label{ak0}\ee
 without loss of
generality because the terms with $a_k$ in Eq.~(\ref{X}) describe a
uniform rectilinear motion of the system at a constant velocity. It
may be eliminated via Lorentz transformation \cite{MilSh}. In the
case $a_k=0$, in particular, relations (\ref{bak}) transform into
 \be
b_0=-\th a_0,\qquad b_3=0,\qquad b_k=\frac{\ell_kN_k}{2\pi}\quad
(k>3),
 \label{bk}\ee
and other equations (\ref{ABort1})\,--\,(\ref{a0Q})
--- correspondingly.

In the case (\ref{ak0}) one can exclude the amplitude factor
(\ref{AA}) from Eqs.~(\ref{a01}), (\ref{a02}) and obtain the
equation
 \be
1+\theta^2-\theta\frac{2S+(h_1+h_2)\,C-h_1h_2C_1S_2}{S_\th}=
\frac1{a_0^2}\sum_{k>3}b_k^2.
 \label{omth}\ee

Relations between the factor $a_0$, speeds of the massive points
$v_i={}$const and other parameters of the system are result from
Eqs.~(\ref{a01})\,--\,(\ref{a0Q}):
 \be a_0
=\frac{m_1Q_1}{\gamma\sqrt{1-v_1^2}}=\frac{m_2Q_2}{\gamma\sqrt{1-v_2^2}},
\qquad A_*^2=\frac{a_0^2\th S_\th}{\om^2},
 \label{a0A}\ee
\vspace{-1ex}
 \be
v_1^2=\th\frac{S-h_2S_1S_2}{S_\th},\qquad
v_2^2=\th\frac{S-h_1S_1S_2}{S_\th}.
 \label{v12}\ee

Values $\om$ and $\th$ are determined from the system (\ref{eqC}),
 and (\ref{omth}).  Solution of the
system (\ref{eqC}), (\ref{omth}) (pairs $\om$, $\th$) form some
countable set. Each pair corresponds to solution (\ref{X})
describing uniform rotation of the closed string with certain
topological type.

To investigate the obtained world surface (\ref{X}) one can consider
its section $t=t_0={}$const --- a ``photograph" of the string
position at time moment $t_0$.

In the case $a_k=b_k=0$  (or for Minkowski space ${\cal M}=R^{1,3}$)
projections of these sections onto $e_1,\,e_2$ plane are closed
curves, composed from segments of a hypocycloid if and only if the
equalities (\ref{eqC}), (\ref{omth}) are fulfilled. This result is
similar to the behavior of rotational motions for the string baryon
model ``triangle" \cite{PRTr}.

Hypocycloid is the curve drawing by a point of a circle (with radius
$r$) rolling inside another fixed circle with larger radius $R$. In
the case of solutions (\ref{X}) uniformly rotating hypocycloidal
segments of the string are joined at non-zero angles in the massive
points. The relation of the mentioned radii
$$
\frac rR=\frac{1-|\theta|}2
$$
is irrational if $m_i\ne0$ and $\th\ne0$.

This hypocycloidal string rotates in the $e_1,\,e_2$ plane at the
angular velocity $\Omega=\om/a_0$, the massive points move at the
speeds (\ref{v12})  along the circles with radii $v_i/\Omega$. There
are also cusps (return points) of the hypocycloid moving at the
speed of light.

The more general solutions (\ref{X}) including the summands with
$b_k\ne0$ (for cyclical coordinates in ${\cal M}$) differ from
mentioned hypocycloidal solutions: their sections $t={}$const are
spatial (not flat) curves, closed because of cyclical nature of
coordinates $x_k$. Their projections onto the plane $e_1,\,e_2$ look
like hypocycloids. But these world surfaces have no peculiarities of
the metric $\dot X^2=X^{'2}=0$.

In the case when the parameter in Eq.~(\ref{tau0}) equals zero
($\th=0$ or $\tau_0=0$) solutions (\ref{X}) describe rotational
motions of $n$ times folded string. It has a form of rotating
rectilinear segment if $b_k=0$ and more complicated form in the case
$b_k\ne0$. These motions are divided into two classes, we shall name
them sa follows: (a) ``linear states'' with masses $m_1$, $m_2$
moving at nonzero velocities $v_1$, $v_2$ at the ends of the
rotating rectilinear segment and (b) ``central states'', if one mass
(or both masses) is placed at the rotational center.

The rotational motions (\ref{X}) in the case $\th\ne0$ we shall name
``hypocycloidal states''.

 There are many topologically different types of linear, central
and hypocycloidal states (\ref{X}). They may be classified with the
number of cusps and the type of intersections of the hypocycloid
following Ref.~\cite{PRTr}. Note that in the considered model
(\ref{S}) the string does not interact with itself in a point of
intersection.

These topological configurations of the rotational states may be
classified by investigation of the massless $m_i\to0$ or
ultrarelativistic $v_i\to1$ limit for fixed $N_k$, $\ell_k$,
$\gamma$, $a_0$. Analysis of equations  (\ref{eqC}),
(\ref{omth})\,--\,(\ref{v12}) shows that in the limit $m_i\to0$ the
values $Q_i$ tend to infinity, values $2\om$ and $2\th\om$ tend to
following integer numbers:
 \be
n_1=\Big|\lim\limits_{m_i\to0}2\om\Big|, \qquad
n_2=\lim\limits_{m_i\to0}2\theta\om.
 \label{n1n2}\ee

Because of the inequality $|\theta|<1$  and condition (\ref{eqC}),
resulting in the equality $(-1)^{n_1}=(-1)^{n_2}$ ($n_1$ and $n_2$
are admissible:
 \be n_1\ge2;\qquad n_2=n_1-2,\;\;
n_1-4,\;\dots\;-(n_1-2).
 \label{n12con}\ee

The number $n_1$ is the number of cusps of the rotating hypocycloid
(including massive points), the number $n_2$ describes the shape of
this curve. For example, values $n_1=5$, $n_2=3$ correspond to a
curvilinear pentagon,  values $n_1=5$, $n_2=1$ --- a curvilinear
star).

 The case $n_2=0$ (this means $\th=0$) includes two classes:
linear and central rotational states. They are joined in the limit
$m_i\to0$.

\bigskip

\centerline {\bf 3. Regge trajectories}
\medskip

The obtained rotational motions of the considered model should be
applied for describing physical manifestations of glueballs, in
particular, their Regge trajectories. For this purpose we calculate
the energy $E$ and angular momentum $J$ for the states (\ref{X}) of
this model. For an arbitrary classic state of the relativistic
string with the action (\ref{S}) carrying pointlike masses they are
determined by the following integrals (Noether currents)
\cite{PRTr,MilSh}:
 \bea &\displaystyle P^\mu=\int\limits_c
p^\mu(\tau,\s)\,d\s +\sum_{i=1}^2p_i^\mu(\tau),&
 \label{Pimp}\\
&\displaystyle {\cal J}^{\mu\nu}=\int\limits_C\Big[X^\mu(\tau,\s)\,
p^\nu(\tau,\s)-X^\nu(\tau,\s)\,p^\mu(\tau,\s)\Big]\,d\s+\sum_{i=1}^2
\Big[x_i^\mu(\tau)\,p_i^\nu(\tau)-x_i^\nu(\tau)\,p_i^\mu(\tau)\Big],&
 \label{Mom}\eea
 where
$p^\mu(\tau,\s)=\gamma\big[(\dot X,X') X^{\pr\mu}-X'{}^2\dot
X^\mu\big]/\sqrt{-g}$ is the canonical string  momentum,
 $x_i^\mu(\tau)=X^\mu\big(\tau,\s_i(\tau)\big)$ and
$p_i^\mu(\tau)=m_i\dot x_i^\mu(\tau)\big/\sqrt{\dot x_i^2(\tau)}$
are coordinates and momentum of the massive points, $C$ is any
closed curve (contour) on the tube-like world surface of the string.
Note that the lines $\tau={}$const on the world surface (\ref{X})
are not closed in the case $\tau_0\ne0$. So we can use the most
suitable lines $\tau-\theta\s={}$const (that is $t={}$const) as the
contour $C$ in integrals (\ref{Pimp}), (\ref{Mom}).

The reparametrization $\tilde\tau=\tau-\theta\s$, $\tilde\s=\s
-\theta\tau$ keeps the orthonormality conditions (\ref{ort}). Under
them $p^\mu(\tau,\s)=\gamma\dot X^\mu(\tau,\s)$.

The square of energy $E^2$ equals the scalar square of the conserved
vector of momentum (\ref{Pimp}): $P^2=P_\mu P^\mu=E^2$. If we
substitute the expression (\ref{X}) in the case (\ref{ak0}) $a_k=0$
into Eq.~(\ref{Pimp}) we'll obtain the following formula for the
momentum:
 \be
P^\mu=\gamma a_0\bigg[2\pi(1-\theta^2)+
\frac1{Q_1}+\frac1{Q_2}\bigg]\,e_0^\mu+ \gamma\th
\sum\limits_{k>3}\ell_k N_k e_k^\mu.
 \label{P}\ee

In the simplest case $N_k=0$ (or for Minkowski space) this
expression takes the form
 \be
P^\mu=Ee_0^\mu,\qquad E=2\pi\gamma a_0(1-\theta^2)+
\frac{m_1}{\sqrt{1-v_1^2}}+\frac{m_2}{\sqrt{1-v_2^2}}.
 \label{P0}\ee

The classical angular momentum (\ref{Mom}) is not conserved value
for the considered system because of anisotropy of the space ${\cal
M}$. Only the components ${\cal J}^{\mu\nu}$ with $\mu,\nu=0,1,2,3$
(relating to the space $R^{1,3}$) are conserved. Among them only
$z$-component of the angular momentum is nonzero:
 \be\begin{array}{c}
{\cal J}^{\mu\nu}= j_3^{\mu\nu}J,\rule[-3mm]{0mm}{1mm}\\
\displaystyle
J=\frac{\gamma}{2\om}\left\{2\pi\bigg[a_0^2(1-\theta^2)-
\sum_{k\ge3}\frac{\ell_k^2N_k^2}{4\pi^2}\bigg]
+a_0^2\bigg(\frac{v_1^2}{Q_1}+\frac{v_2^2}{Q_2}\bigg) \right\}.
 \end{array} \label{J}\ee
 Here $j_3^{\mu\nu}=e_1^\mu e_2^\nu-e_1^\nu
e_2^\mu=e^\mu\acute e^\nu- e^\nu\acute e^\mu.$

For given cyclical numbers $N_k$ the states (\ref{X}) are determined
by the parameters $a_0$, $\om$, $\th$. If the values $m_i$, $\gamma$
and the topological type of the rotational state (\ref{X}) are fixed
we obtain the one-parameter set of motions with different values $E$
and $J$. These states lay at quasilinear Regge trajectories.

The mentioned Regge trajectories are nonlinear for small $E$ and
tend to linear if $E\to\infty$. Their slope in this limit depends on
the values $N_k$ and the fixed topological type.

The ultrarelativistic limit $E\to\infty$ corresponds to $v_i\to1-0$
(except for central states) and for values $\om$ and $\th$ --- to
the limits (\ref{n1n2}). Substituting into Eqs.~(\ref{eqC}),
(\ref{omth}), (\ref{v12}),  (\ref{P}),  (\ref{J}) asymptotic
relations with small values $\ep_1=\sqrt{1-v_1^2}$,
$\ep_2=\sqrt{1-v_2^2}$, $2\om=n_1-\ep_\om$, $n_1\th=n_2-\ep_\th$, we
obtain in the limit $J\to\infty$, $E\to\infty$ the following
asymptotic relation between these values for fixed type
($n_1,\,n_2,\,b_k$) of the state:
 \be
J\simeq\al'E^2+\al_1E^{1/2}+\al_0, \qquad E\to\infty,
 \label{JElim}\ee
 where
 \be
 \al'=\frac1{2\pi\gamma}\,\frac{n_1}{n_1^2-n_2^2},
 \label{slope}\ee
$$\al_1=-\frac{\sqrt2\,n_1(m_1^{3/2}+m_2^{3/2})}{3\sqrt\pi\gamma(n_1^2-n_2^2)^{3/4}},\qquad
\al_0=-\frac{2\pi\gamma(n_1^2-2n_2^2)}{n_1(n_1^2-n_2^2)}\sum\limits_{k>3}b_k^2.$$

This dependence is close to linear one (\ref{Reggm}), but the slope
$\al'$  (\ref{slope}) for this system differs from the Nambu value
$\al'=1/(2\pi\gamma)$ by the factor $\chi=n_1/(n_1^2-n_2^2)$. The
maximal slope with the factor $\chi=1/2$ corresponds to $n_1=2$,
$n_2=0$ that is to the linear state with two masses at the ends,
connected two strings without singularities. For the ``triangle''
states  $n_1=3$, $n_2=1$ this factor is $\chi=3/8$ and these states
(and also states with $n_1=4$, $n_2=2$, $\chi=1/3$) are more
suitable candidates for describing the glueball trajectory
(\ref{ReggPom}).

\bigskip

\newpage

\centerline {\bf Conclusion}
\medskip

The obtained rotational states (\ref{X}) of the  closed string with
two point-like masses are divided in a set of different topological
classes, describing by the integer parameters (\ref{n1n2}). The
states from different classes generate the wide spectrum of
quasilinear Regge trajectories  (\ref{P}), (\ref{J}) with different
slopes (\ref{J}) in the limit (\ref{JElim}) of large energies. There
are some classes of these states suitable for describing the pomeron
(glueball) trajectory (\ref{ReggPom}).

The considered model needs further development, in particular,
quantization or quantum corrections. These corrections are to be
significant for calculation of the intercept $\al_0$.

\medskip

\centerline{\bf Acknowledgment}

\medskip

The author is grateful to Russian foundation of basic research for
support.

\end{document}